\documentclass{article}
\usepackage{cite}
\usepackage{graphicx}
\usepackage{dcolumn}

\begin{document}

\date{}
\title{Comment on: ``Relativistic quantum dynamics of a charged particle in cosmic
string spacetime in the presence of magnetic field and scalar potential'' .
Eur. Phys. J. C (2012) 72:2051}
\author{Francisco M. Fern\'{a}ndez\thanks{%
fernande@quimica.unlp.edu.ar} \\
%EndAName
INIFTA, DQT, Sucursal 4, C.C 16, \\
1900 La Plata, Argentina}
\maketitle

\begin{abstract}
We analyze the results of a paper on ``Relativistic quantum dynamics of a
charged particle in cosmic string spacetime in the presence of magnetic
field and scalar potential''. We show that the authors did not obtain the
spectrum of the eigenvalue equation but only one eigenvalue for a specific
relationship between model parameters. In particular, the existence of
allowed cyclotron frequencies conjectured by the authors is a mere artifact
of the truncation condition used to obtain exact solutions to the radial
eigenvalue equation.
\end{abstract}

\section{Introduction}

\label{sec:intro}

In a paper published in this journal Figueiredo Medeiros and Becerra de Mello%
\cite{FB12} analyze the relativistic quantum motion of charged spin-0 and
spin-$\frac{1}{2}$ particles in the presence of a uniform magnetic field and
scalar potentials in the cosmic string spacetime. They derive an eigenvalue
equation for the radial coordinate and solve it exactly by means of the
Frobenius method. This approach leads to a three-term recurrence relation
that enables the authors to truncate the series and obtain eigenfunctions
with polynomial factors. They claim to have obtained the energy spectrum of
the model and the truncation condition requires that the cyclotron frequency
or other model parameters depend on the quantum numbers. In this Comment we
analyze the effect of the truncation condition used by the authors on the
physical conclusions that they derive in their paper. In section~\ref
{sec:TTRR} we apply the Frobenius method, derive a three-term recurrence
relation for the coefficients and analyze the results obtained in this way.
Finally, in section~\ref{sec:conclusions} we summarize the main results and
draw conclusions.

\section{The truncation method}

\label{sec:TTRR}

It is not our purpose to discuss the validity of the models but the way in
which the authors solve the eigenvalue equation. For this reason we do not
show the main equations displayed in their paper and restrict ourselves to
what we consider relevant. We just mention that the authors state that they
choose natural units such that $\hbar =c=G=1$. A rigorous way of deriving
dimensionless equations, as well as the choice of natural units, is reviewed
in a recent pedagogical paper where we criticize such an unclear way of
introducing them\cite{F20}.

Some of the authors' eigenvalue equations are particular cases of
\begin{eqnarray}
\hat{L}R &=&WR,  \nonumber \\
\hat{L} &\equiv &-\frac{d^{2}}{d\xi ^{2}}-\frac{1}{\xi }\frac{d}{d\xi }+%
\frac{\gamma ^{2}}{\xi ^{2}}-\frac{a}{\xi }+b\xi +\xi ^{2},
\label{eq:eigen_eq_R}
\end{eqnarray}
where $\gamma $, $a$ and $b$ are real numbers and $\gamma $ depends on the
rotational quantum number $m=0,\pm 1,\pm 2,\ldots $. By means of the ansatz
\begin{equation}
R(\xi )=\xi ^{|\gamma |}e^{-\frac{b\xi }{2}-\frac{\xi ^{2}}{2}}P(\xi
),\,P(\xi )=\sum_{j=0}^{\infty }c_{j}\xi ^{j},  \label{eq:R_series}
\end{equation}
we obtain a three-term recurrence relation for the coefficients $c_{j}$:
\begin{eqnarray}
c_{j+2} &=&\frac{b\left( 2\gamma +2j+3\right) -2a}{2\left( j+2\right) \left(
2\gamma +j+2\right) }c_{j+1}+\frac{4\left( 2\gamma +2j-W+2\right) -b^{2}}{%
4\left( j+2\right) \left( 2\gamma +j+2\right) }c_{j},  \nonumber \\
j &=&-1,0,1,\ldots ,\;c_{-1}=0,\;c_{0}=1.  \label{eq:rec_rel_gen}
\end{eqnarray}

In order to obtain ``a special kind of exact solutions representing bound
states'' the authors require the termination conditions
\begin{equation}
W=W_{m}^{(n)}=2\left( \gamma +n+1\right) -\frac{b^{2}}{4},\,c_{n+1}=0,%
\,n=0,1,\ldots .  \label{eq:trunc_cond}
\end{equation}
Clearly, under such conditions $c_{j}=0$ for all $j>n$ and $P(\xi
)=P_{m}^{(n)}$ reduces to a polynomial of degree $n$. In this way, they
obtain analytical expressions for the eigenvalues $W_{m}^{(n)}$ and the
radial eigenfunctions $R_{m}^{(n)}(\xi ).$ For the sake of clarity and
generality we will use $\gamma $ instead of $m$ as an effective quantum
number.

For example, when $n=0$ we have
\begin{equation}
a_{0,\gamma }=\frac{b\left( 2\gamma +1\right) }{2},\;W_{\gamma
}^{(0)}=2\left( \gamma +1\right) -\frac{b^{2}}{4}.  \label{eq:W,a,n=0}
\end{equation}
When $n=1$ there are two solutions for $a$
\begin{eqnarray}
W_{\gamma }^{(1)} &=&2\left( \gamma +2\right) -\frac{b^{2}}{4},\,a_{1,\gamma
}^{(1)}=\frac{2b\left( \gamma +1\right) -\sqrt{b^{2}+8\left( 2\gamma
+1\right) }}{2},  \nonumber \\
a_{1,\gamma }^{(2)} &=&\frac{2b\left( \gamma +1\right) +\sqrt{b^{2}+8\left(
2\gamma +1\right) }}{2},  \label{eq:W,a,n=1}
\end{eqnarray}
or, alternatively,
\begin{eqnarray}
b_{1,\gamma }^{(1)} &=&\frac{2\left[ 2a\left( \gamma +1\right) -\sqrt{%
a^{2}+2\left( 2\gamma +3\right) \left( 2\gamma +1\right) ^{2}}\right] }{%
\left( 2\gamma +1\right) \left( 2\gamma +3\right) },  \nonumber \\
b_{1,\gamma }^{(2)} &=&\frac{2\left[ 2a\left( \gamma +1\right) +\sqrt{%
a^{2}+2\left( 2\gamma +3\right) \left( 2\gamma +1\right) ^{2}}\right] }{%
\left( 2\gamma +1\right) \left( 2\gamma +3\right) }.  \label{eq:b,n=1}
\end{eqnarray}
When $n=2$ we obtain a cubic equation for either $a$ or $b$, for example,

\begin{eqnarray}
W_{\gamma }^{(2)} &=&2\left( \gamma +3\right) -\frac{b^{2}}{4},  \nonumber \\
&&4a^{3}-6a^{2}b\left( 2\gamma +3\right) +a\left( b^{2}\left( 12\gamma
^{2}+36\gamma +23\right) -16\left( 4\gamma +3\right) \right)  \nonumber \\
&&-\frac{b\left( 2\gamma +1\right) \left( b^{2}\left( 2\gamma +3\right)
\left( 2\gamma +5\right) -16\left( 4\gamma +7\right) \right) }{2}=0,
\label{eq:a,b,n=2}
\end{eqnarray}
from which we obtain either $a_{2,\gamma }(b)$ or $b_{2,\gamma }(a)$; for
example, $a_{2,\gamma }^{(1)}(b)$, $a_{2,\gamma }^{(2)}(b)$, $a_{2,\gamma
}^{(3)}(b)$. In the general case we will have $n+1$ curves of the form $%
a_{n,\gamma }^{(i)}(b)$, $i=1,2,\ldots ,n+1$, labelled in such a way that $%
a_{n,\gamma }^{(i)}(b)<a_{n,\gamma }^{(i+1)}(b)$. It can be proved that all
these roots are real\cite{CDW00,AF20}

It is obvious to anybody familiar with conditionally solvable (or
quasi-solvable) quantum-mechanical models\cite{CDW00,AF20} (and references
therein) that the approach just described does not produce all the
eigenvalues of the operator $\hat{L}$ for a given set of values of $\gamma $%
, $a$ and $b$ but only those states with a polynomial factor $P_{\gamma
}^{(n)}(\xi )$. Each of the particular eigenvalues $W_{\gamma }^{(n)}$, $%
n=1,2,\ldots $ corresponds to a set of particular curves $a_{n,\gamma
}^{(i)}(b)$ in the plane $a-b$ of physical model parameters. On the other
hand, it is obvious that the eigenvalue equation (\ref{eq:eigen_eq_R})
supports an infinite set of eigenvalues $W_{\nu ,\gamma }(a,b)$, $\nu
=0,1,2,\ldots $ for each set of real values of $a$, $b$ and $\gamma $. The
condition that determines these allowed values of $W$ is that the
corresponding radial eigenfunctions $R(\xi )$ are square integrable
\begin{equation}
\int_{0}^{\infty }\left| R(\xi )\right| ^{2}\xi \,d\xi <\infty ,
\label{eq:bound-state_def_xi}
\end{equation}
as shown in any textbook on quantum mechanics\cite{LL65,CDL77}. Notice that $%
\nu $ is the actual radial quantum number (that labels the eigenvalues in
increasing order of magnitude and the number of nodes of the corresponding
radial eigenfunctions), whereas $n$ is just a positive integer that labels
some particular solutions with a polynomial factor $P_{\gamma }^{(n)}(\xi )$%
. In other words: $n$ is a fictitious quantum number given by the truncation
condition (\ref{eq:trunc_cond}). More precisely, $W_{\gamma }^{(n)}$ is an
eigenvalue of a given operator $\hat{L}_{n,\gamma }$ whereas $W_{\gamma
^{\prime }}^{(n^{\prime })}$ is an eigenvalue of a different linear operator
$\hat{L}_{n^{\prime },\gamma ^{\prime }}$; for this reason one does not
obtain the spectrum of a given quantum-mechanical system by means of the
truncation condition (\ref{eq:trunc_cond}). The situation is even worse if
one takes into consideration that $\hat{L}_{n,\gamma }$ actually means $\hat{%
L}_{n,\gamma ,i}$, $i=1,2,\ldots ,n+1$.

It should be obvious to everybody that the eigenvalue equation (\ref
{eq:eigen_eq_R}) supports bound states for all real values of $a$ and $b$
and that the truncation condition (\ref{eq:trunc_cond}) only yields some
particular solutions for some particular model operators $\hat{L}$. Besides,
according to the Hellmann-Feynman theorem\cite{CDL77,P68} (and references
therein) the true eigenvalues $W_{\nu ,\gamma }(a,b)$ of equation (\ref
{eq:eigen_eq_R}) are decreasing functions of $a$ and increasing functions of
$b$%
\begin{equation}
\frac{\partial W}{\partial a}=-\left\langle \frac{1}{\xi }\right\rangle ,\,%
\frac{\partial W}{\partial b}=\left\langle \xi \right\rangle .
\label{eq:HFT}
\end{equation}
Therefore, for a given value of $b$ and sufficiently large values of $a$ we
expect negative values of $W$ that the truncation condition fails to
predict. It is not difficult to prove, from straightforward scaling\cite{F20}%
, that
\begin{equation}
\lim_{a\rightarrow \infty }\frac{W_{\nu ,\gamma }}{a^{2}}=-\frac{1}{\left(
2\nu +2\gamma +1\right) ^{2}},  \label{eq:W_asympt}
\end{equation}
for any given value of $b$. What is more, from equation (\ref{eq:HFT}) we
can conjecture that the pairs $\left[ a_{n,\gamma }^{(i)}(b),W_{\gamma
}^{(n)}\right] $, $i=1,2,\ldots ,n+1$ are points on the curves $%
W_{i-1,\gamma }(a,b)$ for a given value of $b$.

The eigenvalue equation (\ref{eq:eigen_eq_R}) cannot be solved exactly in
the general case (contrary to what the authors appear to believe). In order
to obtain sufficiently accurate eigenvalues of the operator $\hat{L}$ we
resort to the reliable Rayleigh-Ritz variational method that is well known
to yield increasingly accurate upper bounds to all the eigenvalues of the
Schr\"{o}dinger equation\cite{P68} (and references therein). For simplicity
we choose the basis set of non-orthogonal functions $\left\{ u_{j}(\xi )=\xi
^{|\gamma |+j}e^{-\frac{\xi ^{2}}{2}},\;j=0,1,\ldots \right\} $. We test the
accuracy of these results by means of the powerful Riccati-Pad\'{e} method%
\cite{FMT89a}.

As a first example, we choose $n=2$, $\gamma =0$ and $b=1$ so that $%
W_{0}^{(2)}=5.75$ for the three models $\left[
a_{2,0}^{(1)}=-1.940551663,b=1\right] $, $\left[
a_{2,0}^{(2)}=1.190016441,b=1\right] $ and $\left[
a_{2,0}^{(3)}=5.250535221,b=1\right] $. The first four eigenvalues for each
of these models are
\begin{eqnarray*}
a_{2,0}^{(1)} &\rightarrow &\left\{
\begin{array}{c}
W_{0,0}=5.750000000 \\
W_{1,0}=9.894040660 \\
W_{2,0}=14.06831985 \\
W_{3,0}=18.24977457
\end{array}
\right. , \\
a_{2,0}^{(2)} &\rightarrow &\left\{
\begin{array}{c}
W_{0,0}=-0.1664353619 \\
W_{1,0}=5.750000000 \\
W_{2,0}=10.52307155 \\
W_{3,0}=15.06421047
\end{array}
\right. , \\
a_{2,0}^{(3)} &\rightarrow &\left\{
\begin{array}{c}
W_{0,0}=-27.32460313 \\
W_{1,0}=-0.5108147276 \\
W_{2,0}=5.750000000 \\
W_{3,0}=10.90599171
\end{array}
\right. .
\end{eqnarray*}
We appreciate that the eigenvalue $W_{0}^{(2)}=5.75$ coming from the
truncation condition (\ref{eq:trunc_cond}) is the lowest eigenvalue of the
first model, the second lowest eigenvalue of the second model and the third
lowest eigenvalue for the third model (in agreement with the conjecture put
forward above). The truncation condition misses all the other eigenvalues
for each of those models and for this reason it cannot provide the spectrum
of the physical model for any set of values of $\gamma $, $a$ and $b$,
contrary to what is suggested by Figueiredo Medeiros and Becerra de Mello%
\cite{FB12}.

In the results shown above we have chosen model parameters on the curves $%
a_{2,0}^{(i)}(b)$. In what follows we consider the case $a=2$, $b=1$ that
does not belong to any curve $a_{n,\gamma }^{(i)}$ (that is to say, it does
not stem from the truncation condition). For this set of model parameters,
the first five eigenvalues are $W_{0,0}=-3.230518994$, $W_{1,0}=4.510929109$%
, $W_{2,0}=9.532275968$, $W_{3,0}=14.19728140$ and $W_{4,0}=18.70978427$. As
said above: there are square-integrable solutions (actual bound states) for
any set of real values of $a$, $b$ and $\gamma $. The obvious conclusion is
that the dependence of the cyclotron frequency $\omega $ or other
parameters, like $\eta _{L}$, on the quantum numbers conjectured by
Figueiredo Medeiros and Becerra de Mello\cite{FB12} is just an artifact of
the truncation condition (\ref{eq:trunc_cond}). Such claims are nonsensical
from a physical point of view.

\begin{figure}[tbp]
\begin{center}
\includegraphics[width=9cm]{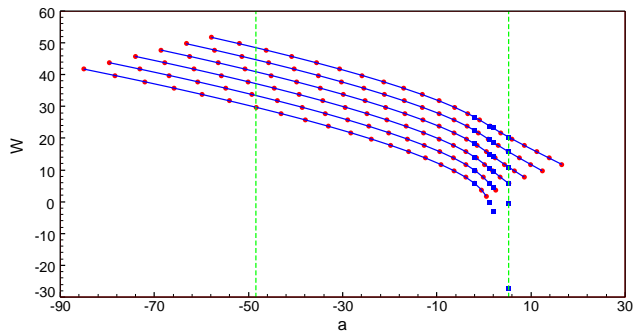}
\end{center}
\caption{Eigenvalues $W_{\nu,0}(a,1)$, $\nu=0,1,2,3,4,5$ obtained from the
truncation condition (red circles) and numerically (blue squares) }
\label{fig:W}
\end{figure}

The red circles in figure~\ref{fig:W} denote some of the eigenvalues $%
W_{0}^{(n,i)}(a,1)$ given by the truncation condition (\ref{eq:trunc_cond})
and the blue lines connect those corresponding to the actual eigenvalues $%
W_{\nu ,0}(a,1)$. Some eigenvalues calculated numerically by the methods
mentioned above are marked by blue squares. The energy spectrum of a model
given by a pair of values of $a$ and $b$ is determined by all the
intersections between a vertical line and the blue ones. Such intersections
meet at most one red circle (left green dashed line, for example). The right
green dashed line cuts the blue lines at some of the eigenvalues calculated
numerically.

\begin{figure}[tbp]
\begin{center}
\includegraphics[width=9cm]{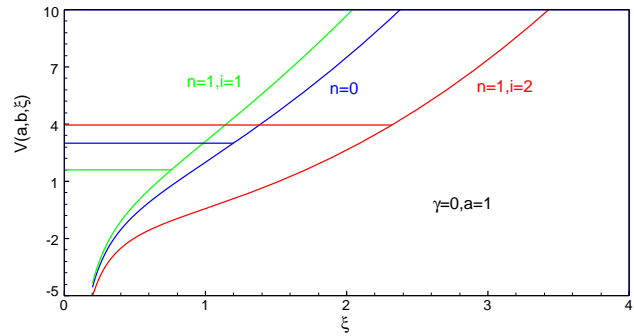}
\end{center}
\caption{Potentials $V(a,b,\xi)=-a/\xi+b \xi+\xi^2$ for $\gamma=0$, $a=1$
and $n=0,1$. The horizontal lines are the corresponding energies $%
W_\gamma^{(n,i)}$}
\label{fig:potsw}
\end{figure}

Figure~\ref{fig:potsw} shows three potentials $V(a,b,\xi )=-a/\xi +b\xi +\xi
^{2}$ for $\gamma =0$, $a=1$ and $b=b_{n,0}^{(i)}$. They are given by $n=0$
and $n=1$, with $i=1,2.$ Three horizontal lines indicate the eigenvalues $%
W_{0}^{(0)}$, $W_{0}^{(1,1)}$ and $W_{0}^{(1,2)}$; their purpose being to
make clearer that the eigenvalues $W_{\gamma }^{(n,i)}$ correspond to
different models $V\left( a,b_{n,\gamma }^{(i)},\xi \right) $ and do not
give the spectrum of a single model.

\section{Conclusions}

\label{sec:conclusions}

The authors make two basic, conceptual errors. The first one is to believe
that the only possible bound states are those given by the truncation
condition (\ref{eq:trunc_cond}) that have polynomial factors $P_{\gamma
}^{(n)}(\xi )$. We have argued above that there are square-integrable
solutions for all real values of $a$ and $b$ and calculated some of them
outside the curves $a_{n,\gamma }^{(i)}(b)$ associated to these polynomials.
The second error, connected with the first one, is the assumption that the
spectrum of the problem is given by the eigenvalues $W_{\gamma }^{(n)}$
stemming from that truncation condition. It is clear that the truncation
condition only provides one energy eigenvalue $W_{\gamma }^{(n)}$ for a
particular set of model parameters given by the curves $a_{n,\gamma }^{(i)}$
discussed above. For this reason, the supposedly necessary dependence of the
model parameters on the quantum numbers does not have mathematical support.
Such unphysical conclusions stem from an arbitrary truncation condition that
only produces particular bound states with no special meaning. We have
illustrated these points by means of some numerical calculations and two
figures that, hopefully, are clear enough to disclose the misunderstanding
about the meaning of the results for this conditionally solvable
quantum-mechanical model.

\end{document}